\documentclass[letterpaper,twocolumn,10pt]{article}
\usepackage{usenix2024_SOUPS}

\usepackage{tikz}
\usepackage{amsmath}
\usepackage{graphicx}  
\newcommand{\ignore}[1]{}

\usepackage{filecontents}
\usepackage{amsmath}
\usepackage{amssymb}  
\usepackage{booktabs}
\usepackage{amssymb}
\usepackage{float} 
\usepackage{caption}

\begin{document}

\date{}

\title{\Large \bf Multiverse Privacy Theory for Contextual Risks in Complex User-AI Interactions}

\def\plainauthor{Author name(s) for PDF metadata. Don't forget to anonymize for submission!}

\author{
{\rm Ece Gumusel}\\
Indiana University Bloomington
} 

\maketitle
\thecopyright

\begin{abstract}
In an era of increasing interaction with artificial intelligence (AI), users face evolving privacy decisions shaped by complex, uncertain factors. This paper introduces Multiverse Privacy Theory, a novel framework in which each privacy decision spawns a parallel universe, representing a distinct potential outcome based on user choices over time. By simulating these universes, this theory provides a foundation for understanding privacy through the lens of contextual integrity, evolving preferences, and probabilistic decision-making. Future work will explore its application using real-world, scenario-based survey data.


\end{abstract}

\section{Introduction}
As artificial intelligence (AI) technologies increasingly rely on vast datasets for training and operation, the need to protect user privacy has become more critical. This concern extends beyond just tracking user behavior \cite{gumusel2025literature, zhang2024s}—it also includes protecting against privacy-targeted malicious attempts and adversarial attacks \cite{oseni2021security}, which are becoming easier to execute \cite{Shayea_Zabil_Habeeb_Khaleel_Albahri_2025, gupta2023chatgpt, chakraborty2023artificial}. 
In addition, despite the growing urgency to address these threats, many developers still have a limited understanding of user privacy expectations and its broader implications \cite{Benk_Wettstein_Schlicker_Wangenheim_Scharowski_2025, hadar2018privacy}. They often view privacy narrowly as a matter of regulatory compliance, focused mainly on safeguarding personally identifiable information (PII) \cite{falcao2024investigating}. However, it is a much more complex issue—one that also involves respecting user expectations, ensuring transparency, and fostering trust and satisfaction in AI systems. Especially in accounting for evolving privacy preferences highlighted in previous literature, the dynamic nature of security and privacy threats, and the diversity of user demographics—such as minors, older adults, individuals with disabilities, LGBTQ+ individuals, or users experiencing mental health challenges—whose privacy concerns, disclosure patterns, and trust dynamics vary significantly from the general population \cite{mcdonald2022privacy, citron2025rethinking, geeng2022like, Frik_Nurgalieva_Bernd_Lee_Schaub_Egelman_2019}. Thus, user privacy decisions should not be static or one-size-fits-all. Instead, they should be evaluated within the context of multiple potential infinite number of scenarios, each representing a different privacy decision under various circumstances. 
To address these risks, notable privacy–utility trade-off models have been developed, aiming to balance data usefulness with effective privacy protection. For example, differential privacy (DP) ensures the risk of identifying any individual in a dataset remains low by analyzing pairs of neighboring databases differing by one record \cite{dwork2014algorithmic}. Statistical Data Disclosure (SDC) techniques vary in privacy strength depending on database size and influence later data analysis. Slavković and Seeman's Statistical Data Privacy (SDP) framework \cite{slavkovic2023statistical} also offers an analysis of data release mechanisms that sanitize outputs based on confidential data while considering broader statistical disclosure risks by extending SDC and DP. 




Contextual Integrity (CI) \cite{Nissenbaum_2009} also provides a critical lens of privacy frameworks by defining privacy as the appropriateness of information flows, emphasizing that privacy norms are context-dependent and shift based on the social and informational settings in which data are used. In other words, privacy is preserved when information flows align with the contextual norms in governing 5-parameters: actors (senders, recipients, and subjects), transmission principles or condition, information type, information norms, and user groups \cite{Nissenbaum_2009}. 
Recent work has also integrated the governing knowledge commons framework (GKC) with CI—forming the unified GKC-CI model—which examines how information flows are shaped by institutional rules, social roles, and shared resources \cite{Shvartzshnaider_Sanfilippo_Apthorpe_2022}. 

However, current privacy models typically operate under static assumptions and struggle to capture the full spectrum of user expectations, CI and/or GKC-CI factors, and system-specific risks. These models often prioritize measurable outcomes—such as data minimization or risk scores—while overlooking less quantifiable but equally critical aspects like user trust, perceived control, and the dynamic interplay between evolving privacy harms, risks, and threats. As a result, a significant gap persists between theoretical privacy protections and practical, user-centered outcomes in real-world AI systems. To be effective in dynamic, data-intensive environments—where privacy expectations are fluid, security risks evolve rapidly, and user demographics are increasingly diverse—these models require further extension. Despite various efforts to address these challenges \cite{Chanenson_Pickering_Apthrope_2025, jia2017contexlot, Kumar_Naik_Devkar_Chetty_Clegg_Vitak_2017, apthorpe2018discovering}, their primarily normative focus often falls short in operationalizing privacy within technical systems that demand continuous adaptation, personalization, and real-time decision-making \cite{zimmer2018addressing}. As such, there is a growing need for integrative approaches that bridge conceptual insights with actionable mechanisms for privacy-aware AI design.

\noindent\textbf{Our Approach.} To overcome these limitations, we propose the Multiverse Privacy Theory (MPT) integrates principles from multiverse \cite{deutsch2010apart, mensky2011phenomenology} and CI theories \cite{Nissenbaum_2009} to better handle complex computational systems and diverse user attributes. The idea behind applying multiverse theory to user privacy decision-making within computational systems revolves around the concept of decision-making across multiple potential time-dependent realities or scenarios, each of which reflects different choices and outcomes based on varying user characteristics (e.g., demographics, awareness, comfort), privacy settings, security measures, trust levels, and contextual harms and risks factors. By integrating the MPT, the systems could explore and predict various possible outcomes, each representing a different universe of potential decisions and consequences. 

\noindent\textbf{Contribution.} With MPT, this paper contributes a novel lens for privacy modeling, grounded in the assumption that user privacy experiences are not singular but manifold. 
The theory offers:

\begin{itemize}
    \item A probabilistic understanding of privacy under uncertainty.
    \item An explainable framework for personalized privacy-related decisions.
    \item A bridge between CI and empirical metrics (e.g., utility, trust, risk).
\end{itemize}

\noindent The ultimate goal is modeling the various potential states of privacy, risk, and trust, and how these states evolve over time to optimize decisions for the user. MPT also aims to balance data governance strategies, personalize privacy controls, and analyze AI system behavior under varying sociotechnical conditions.




\section{The Multiverse Privacy Theory}

To formalize the multiverse-inspired approach to privacy decision-making over time, we define a model in which each user action leads to a set of possible universes reflects a different scenario shaped by both user and system context and choice. 

\noindent\textbf{Definition.} Let at each time step \( t \), a set of possible universes \( \mathcal{U}_t = \{ U_t^1, U_t^2, \dots, U_t^n \} \) arises from the user's selected privacy action \( a_t \in \mathcal{A}_t \), where \( \mathcal{A}_t \) denotes the available privacy actions at that time. Each universe reflects how varying contextual factors and the chosen privacy action influence outcomes. The probability of a universe \( U_t^i \) occurring, given privacy action \( a_t \) and context \( C_t \), is modeled as:

\begin{equation}
P(U_t^i \mid a_t, C_t)
\end{equation}

The utility of taking a privacy action \( a_t \) in context \( C_t \) is defined as:

\begin{equation}
\text{UI}(a_t, C_t) = \alpha \cdot \rho_t + \beta \cdot S_t - \gamma \cdot R_t + \delta \cdot T_t + \zeta \cdot g(D_t) + \theta \cdot \text{CI}(a_t, C_t)
\end{equation}

where:
\begin{itemize}
    \item \( \rho_t \): User’s privacy preference at time \( t \)
    \item \( S_t \): Security level of the system/environment at time \( t \)
    \item \( R_t \): Risk level faced by the user in the current context
    \item \( T_t \): User’s trust in the system at time \( t \)
    \item \( g(D_t) \): Influence of demographic attributes such as age, sexual orientation, political affiliation, and familiarity with technology, etc.
    \item \( \text{CI}(a_t, C_t) \): CI score of the privacy action in the current context
\end{itemize}

\noindent Coefficients \( \alpha, \beta, \gamma, \delta, \zeta, \theta \) are tunable weights representing the relative importance of each component in the utility function. To choose the optimal privacy action \( a_t^* \), the system seeks to maximize the expected utility over all possible universe outcomes:

\begin{equation}
a_t^* = \arg\max_{a_t \in \mathcal{A}_t} \; \mathbb{E}[\text{UI}(a_t, C_t)]
\end{equation}

\noindent where the expected utility is defined as:

\begin{equation}
\mathbb{E}[\text{UI}(a_t, C_t)] = \sum_{i=1}^{n} P(U_t^i \mid a_t, C_t) \cdot \text{UI}(a_t, C_t, U_t^i)
\end{equation}

\noindent Here, \( \text{UI}(a_t, C_t, U_t^i) \) denotes the utility of action \( a_t \) considering the outcome in universe \( U_t^i \), possibly adjusted for universe-specific consequences. In addition, MPT introduces a recursive component to account for long-term consequences of privacy actions. The value function at time \( t \) is given by:

\begin{equation}
V_t = \mathbb{E}[\text{UI}(a_t, C_t)] + \lambda \cdot V_{t+1}
\end{equation}

\noindent where \( \lambda \in [0,1] \) is a discount factor representing the weight given to future utility relative to the present. The recursive formulation enables the system to adapt and optimize privacy actions over time as user preferences and contextual factors evolve.


\noindent\textbf{Example.} To illustrate MPT, we implemented a Monte Carlo-style simulation across 5 distinct universes, each representing a plausible configuration of user preferences and contextual factors. Each universe $U_i$ was simulated over 10 time steps. At each time step $t$, values were randomly generated to represent privacy preferences, security level, contextual risk, trust, and demographic sensitivity. Each privacy decision at a given time step resulted in multiple possible outcomes, reflecting alternative universes influenced by these factors. A utility value was computed for each outcome using a weighted combination of the input variables, with weights set to 1.0 for privacy preference~(\(\alpha\)), 0.8 for security level~(\(\beta\)), -0.9 for contextual risk~(\(\gamma\)), 0.6 for trust~(\(\delta\)), and 0.5 for demographic sensitivity~(\(\zeta\)). Additionally, to quantify how well a privacy decision aligns with favorable outcomes in the current context, we computed a CI score for each privacy action at time \( t \) as:

\vspace{-11pt}
\begin{equation}
CI_i(t) = \frac{P_i(t) + S_i(t) + T_i(t) + D_i(t)}{1 + R_i(t)}
\end{equation}
\vspace{-11pt}

where \( P_i(t) \), \( S_i(t) \), \( T_i(t) \), \( D_i(t) \), and \( R_i(t) \) represent normalized values for privacy preference, security level, trust, demographic sensitivity, and risk, respectively. The CI score increases with higher privacy preference, security, trust, and demographics, and decreases with increasing contextual risk.

At each time step, the system evaluates all possible privacy actions by calculating the expected utility over all universes and selects the action that maximizes this expected utility. Utility values were tracked over time and analyzed across three contextual risk bands: low, moderate, and high. Figure~\ref{fig:multiverserisk} illustrates the evolution of utility across the 10 time steps, showing how contextual risk influences privacy-related decisions.

\begin{figure}[H]
    \vspace{-11pt}
    \centering
    \includegraphics[width=1\columnwidth, keepaspectratio]{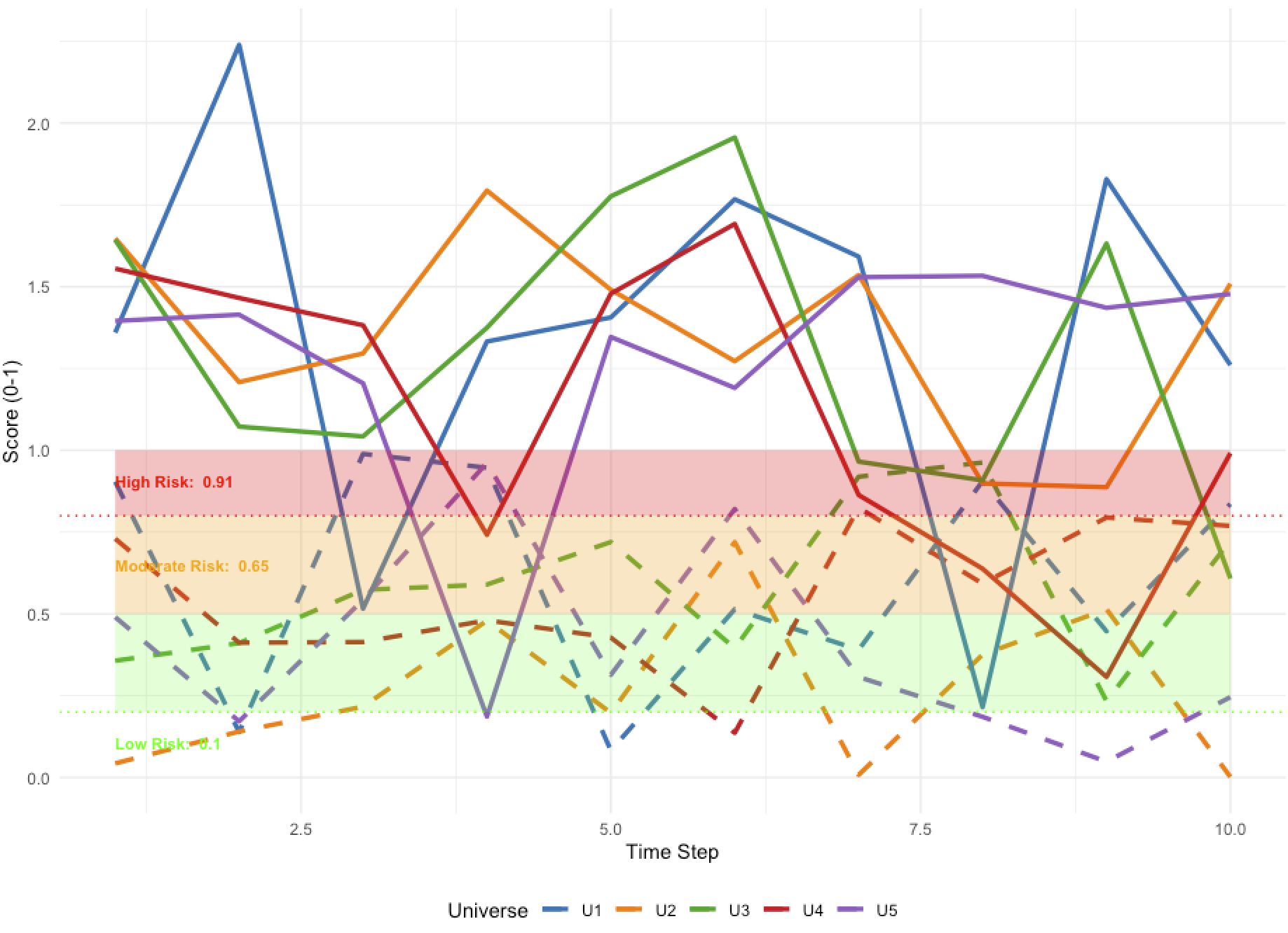}
    \caption{\small Simulation showing how user privacy utility evolves across multiple universes and three contextual risk bands.}
    \label{fig:multiverserisk}
    \vspace{-11pt}
\end{figure}

\paragraph{Hypotheses Testing.} We also tested the relationships between different factors and the utility score using Pearson's correlation. The results are summarized as follows:

\begin{itemize}
    \item[\textbf{H1:}] \textbf{Privacy preferences significantly affect utility.}  
    A strong positive correlation was found between privacy preference and utility (\( r = 0.6618 \), \( p < 0.001 \)), indicating that users who value privacy more tend to derive greater utility from privacy-related decisions. This result supports the hypothesis and is statistically significant.

    \item[\textbf{H2:}] \textbf{Higher contextual risk reduces privacy utility.}  
    There is a strong negative correlation between contextual risk and utility (\( r = -0.6078 \), \( p < 0.001 \)). This indicates that as contextual risk increases, perceived utility decreases. The result is statistically significant and supports the model’s assumptions.

    \item[\textbf{H3:}] \textbf{Trust increases privacy utility.}  
    A moderate positive correlation was observed between trust and utility (\( r = 0.4407 \), \( p < 0.01 \)). This result is statistically significant and supports the hypothesis that trust positively influences privacy-related satisfaction.

    \item[\textbf{H4:}] \textbf{Security level does not significantly affect privacy utility.}  
    The correlation between security level and utility was weak and not statistically significant (\( r = 0.1581 \), \( p = 0.2728 \)). This suggests that, in this model, security level alone may not directly impact perceived utility. One possible explanation is that users’ perceived utility depends more strongly on direct privacy preferences, risk, and trust rather than on abstract or technical security metrics. It is also possible that users may lack sufficient understanding of security levels. However, this needs further exploration.

    \item[\textbf{H5:}] \textbf{CI strongly enhances privacy utility.}  
    A very strong positive correlation was found between CI and utility (\( r = 0.8129 \), \( p < 0.001 \)), supporting the hypothesis that privacy decisions aligned with contextual expectations yield high utility.
\end{itemize}

\begin{table*}[ht]
\centering
\caption{Summary of Hypothesis Testing Results}
\begin{tabular}{@{}clccc@{}}
\toprule
\textbf{Hypothesis} & \textbf{MPT Attributes} & \textbf{r} & \textbf{p-value} & \textbf{95\% CI} \\ \midrule
\textbf{H1} & Privacy Preference – Utility & \textbf{0.6618} & \textbf{1.668e-07} & [0.4700, 0.7939] \\
\textbf{H2} & Contextual Risk – Utility & \textbf{-0.6078} & \textbf{2.855e-06} & [-0.7579, -0.3965] \\
\textbf{H3} & Trust – Utility & \textbf{0.4407} & \textbf{0.0014} & [0.1871, 0.6380] \\
\textbf{H4} & Security Level – Utility & 0.1581 & 0.2728 & [-0.1258, 0.4181] \\
\textbf{H5} & CI – Utility & \textbf{0.8129} & \textbf{7.544e-13} & [0.6908, 0.8899] \\
\bottomrule
\end{tabular}
\end{table*}

By considering these factors, the system can simulate multiple potential scenarios across a wider range of privacy preferences, security concerns, and contextual influences. In other words, MPT can address evaluating these different universes by incorporating user demographics into each simulation, ultimately determining the optimal privacy decisions for each individual user. 
These decisions are made dynamically and adjust as the user's preferences evolve over time, leading to a personalized privacy experience that is adaptive and context aware. In addition, in this model, a system can continuously learn and adjust its decisions based on changing user demographics, ensuring that the best privacy practices are applied in ways that align with the user's preferences, context, and trust level. This process is inherently probabilistic, as the system must choose between several possible universes of future interactions. By evaluating multiple potential universes, the system can mitigate risks of data exposure and maintain a balance between user privacy and security across various contexts. 

\ignore{
\section{Simulation Example and Evaluation}
To illustrate MPT, we implemented a Monte Carlo-style simulation across $n=5$ distinct universes, each representing a plausible configuration of user preferences and contextual factors. Each universe $U_i$ was simulated over $T=10$ time steps. At each time step $t$, the following variables were randomly drawn from uniform distributions to represent key components of privacy decision-making:

\begin{itemize}
    \item $P_i(t) \sim \mathcal{U}(0, 1)$: Privacy preference
    \item $S_i(t) \sim \mathcal{U}(0.5, 1)$: Security level
    \item $R_i(t) \sim \mathcal{U}(0, 1)$: Contextual risk
    \item $T_i(t) \sim \mathcal{U}(0, 1)$: Trust level
    \item $D_i(t) \sim \mathcal{U}(0.2, 0.9)$: Demographic sensitivity
\end{itemize}

Each privacy decision at time $t$ ($a_t \in \mathcal{A}_t$) leads to a set of possible outcomes represented by a set of universes, $U_t^{1}, U_t^{2}, \dots, U_t^{n}$. These universes capture different possible future states based on user preferences, security, and contextual risks. We computed the utility function $U_i(t)$ for each universe and time step:

\begin{equation}
U_i(t) = \alpha P_i(t) + \beta S_i(t) - \gamma R_i(t) + \delta T_i(t) + \zeta D_i(t)
\end{equation}

where the weights $\alpha$, $\beta$, $\gamma$, $\delta$, and $\zeta$ were empirically set as: $\alpha=1$, $\beta=0.8$, $\gamma=0.9$, $\delta=0.6$, and $\zeta=0.5$. Additionally, the CI score for each privacy action at time $t$ was computed as:

\begin{equation}
CI_i(t) = \frac{P_i(t) + S_i(t) + T_i(t) + D_i(t)}{1 + R_i(t)}
\end{equation}

At each time step $t$, the system evaluates the possible privacy actions and chooses the one that maximizes expected utility, considering all possible universes that arise from that action. This is a probabilistic decision-making framework, with each universe representing an alternative reality shaped by the user's privacy choice at that moment. We initially compute the expected utility $\mathbb{E}[U_i(t)]$ over all possible universes as:

\begin{equation}
\mathbb{E}[U_i(t)] = \sum_{j=1}^{n} P(U_t^j \mid a_t, C_t) \cdot U_i(t, U_t^j)
\end{equation}

The simulation plots how the utility values for each action evolve over time, considering the recursive nature of the decision-making process. Values for each universe and time step were plotted over time, and summary statistics such as average utility per universe as well as utility per contextual risk band (high, moderate, low), were calculated. Figure \ref{fig:multiverserisk} visualizes the progression of privacy utility over 10 time steps, highlighting how contextual risk influences privacy-related decisions.

\begin{figure}[H]
     \vspace{-20pt} 
    \centering
    \includegraphics[width=1\columnwidth, keepaspectratio]{Screenshot 2025-05-01 at 3.29.14 PM.png}
    \caption{\small Simulation of how user privacy utility evolves across each multiple universes and contextual three risk bands.}
    \label{fig:multiverserisk}
    \vspace{-20pt} 
\end{figure}

\paragraph{Hypotheses Testing.} The following hypotheses were then tested using Pearson's product-moment correlation to explore the relationships between different factors and the utility score. The results reveal several noteworthy insights:

\begin{itemize}
    \item[\textbf{H1:}] \textbf{Privacy preferences significantly affect utility.}  
    A strong positive correlation was found between privacy preference and utility (\( r = 0.6618 \), \( p < 0.001 \)), indicating that users who value privacy more tend to derive greater utility from privacy-related decisions. This result supports the hypothesis and is statistically significant.

    \item[\textbf{H2:}] \textbf{Higher contextual risk reduces privacy utility.}  
    There is a strong negative correlation between contextual risk and utility (\( r = -0.6078 \), \( p < 0.001 \)). This indicates that as contextual risk increases, perceived utility decreases. The result is statistically significant and supports the model’s assumptions.

    \item[\textbf{H3:}] \textbf{Trust increases privacy utility.}  
    A moderate positive correlation was observed between trust and utility (\( r = 0.4407 \), \( p < 0.01 \)). This result is statistically significant and supports the hypothesis that trust positively influences privacy-related satisfaction.

    \item[\textbf{H4:}] \textbf{Security level does not significantly affect privacy utility.}  
    The correlation between security level and utility was weak and not statistically significant (\( r = 0.1581 \), \( p = 0.2728 \)). This suggests that, in this model, security level alone may not directly impact perceived utility.

    \item[\textbf{H5:}] \textbf{CI strongly enhances privacy utility.}  
    A very strong positive correlation was found between CI and utility (\( r = 0.8129 \), \( p < 0.001 \)), supporting the hypothesis that privacy decisions aligned with contextual expectations yield high utility.
\end{itemize}

\begin{table*}[ht]
\centering
\caption{Summary of Hypothesis Testing Results}
\begin{tabular}{@{}clccc@{}}
\toprule
\textbf{Hypothesis} & \textbf{MPT Attributes} & \textbf{r} & \textbf{p-value} & \textbf{95\% CI} \\ \midrule
\textbf{H1} & Privacy Preference – Utility & \textbf{0.6618} & \textbf{1.668e-07} & [0.4700, 0.7939] \\
\textbf{H2} & Contextual Risk – Utility & \textbf{-0.6078} & \textbf{2.855e-06} & [-0.7579, -0.3965] \\
\textbf{H3} & Trust – Utility & \textbf{0.4407} & \textbf{0.0014} & [0.1871, 0.6380] \\
\textbf{H4} & Security Level – Utility & 0.1581 & 0.2728 & [-0.1258, 0.4181] \\
\textbf{H5} & CI – Utility & \textbf{0.8129} & \textbf{7.544e-13} & [0.6908, 0.8899] \\
\bottomrule
\end{tabular}
\end{table*}

In this simplified simulation, we abstracted the infinite universes and recursive decision-making aspects of MPT. Instead, we focused on modeling how utility varies under diverse contexts and user traits, and evaluate whether certain privacy configurations consistently yield higher utility. These results provide empirical support for MPT, initially validating its utility as a framework for understanding privacy decisions across multiple potential universes. 
}

\section{Conclusion and Future Work}
This paper introduces a novel MPT approach to privacy decision-making, considering the dynamic, evolving nature of privacy risks and user preferences. By simulating an infinite set of universes and using a time-dependent recursive model, AI systems can continuously optimize privacy settings in response to changing contexts. 

Future work will explore MPT to test it to the real-world applicability of the model with practical deployment constraints and user expectations. We will explore (1) how to handle real-time threats, (2) refine the model with user demographics and contextual factors, and (3) incorporate regulatory compliance changes. To do this, we will conduct multiple survey design studies including demographic and scenario-based variations to reasonably and statistically evaluate infinite universes with real-world AI user populations. These scenarios will incorporate persona-specific risk models and adaptive utility functions.


\bibliographystyle{plain}
\bibliography{usenix2024_soups_latex-template/usenix2024_SOUPS}

\end{document}